\documentclass[twocolumn,showpacs]{revtex4}
\usepackage{graphicx}
\usepackage{dcolumn}

\begin{document}


\title{Proton-neutron pairing energies in $N=Z$ nuclei at finite temperature}

\author{K. Kaneko$^{1}$ and M. Hasegawa$^{2}$}
\affiliation{
$^{1}$Department of Physics, Kyushu Sangyo University, Fukuoka 813-8503, Japan \\
$^{2}$Laboratory of Physics, Fukuoka Dental College, Fukuoka 814-0193, Japan 
}

\date{\today}

\begin{abstract}
Thermal behavior of isoscalar ($\tau$=0) and isovector ($\tau$=1) 
proton-neutron ({\it pn}) pairing energies
 at finite temperature are investigated by the shell model calculations.
These {\it pn} pairing energies can be estimated by
 double differences of ``thermal" energies 
which are extended from the double differences of binding energies
 as the indicators of {\it pn} pairing energies at zero temperature. 
We found that the delicate balance between isoscalar 
and isovector {\it pn} pairing energies at zero temperature 
disappears at finite temperature. 
When temperature rises, while the isovector {\it pn} pairing energy decreases, 
the isoscalar {\it pn} pairing energy rather increases.
We discuss also the symmetry energy at finite temperature. 
\end{abstract}


\maketitle

The proton-neutron ({\it pn}) pairing energies have become one of hot topics in the study 
of the nuclear structure for proton-rich nuclei. In particular, interests are increasing 
in studying isovector ($\tau$=1) and isoscalar ($\tau$=0) {\it pn} pairing energies 
in medium mass $N=Z$ nuclei produced at the radioactive nuclear beam facilities. The study 
of {\it pn} pairing energies is also important in the astrophysical context. 
These nuclei lie along the explosive rp-process nucleosynthesis path and the nuclear 
properties such as masses, halflives, and isomers have a strong influence on modeling the 
rp-process and identifying possible nucleosynthesis sites. 
Odd-odd $N=Z$ nuclei are an ideal experimental laboratory for the study of $pn$ pairing 
energies. It is well known that the lowest $\tau=0$ and $\tau=1$ states
compete for the ground state changing the sign of the energy 
difference $E_{\tau=1}-E_{\tau=0}$ in odd-odd $N=Z$ nuclei, while all even-even $N=Z$ nuclei 
have the $\tau=0$ ground states. Several authors 
\cite{Vogel,Janecke1,Zeldes,Martinez,Macc,Janecke2,Frauendorf} already pointed out that this 
degeneracy in odd-odd $N=Z$ nuclei reflects the delicate balance between the symmetry energy 
and the like-nucleon neutron-neutron ({\it nn}) (or proton-proton ({\it pp})) pairing energy. 
On the other hand, it has recently been shown that this degeneracy is attributed to 
competition between the isoscalar and isovector pairing energies \cite{Satula,kaneko1,hasegawa}. 
\begin{figure}[b]
\includegraphics[width=8cm,height=10cm]{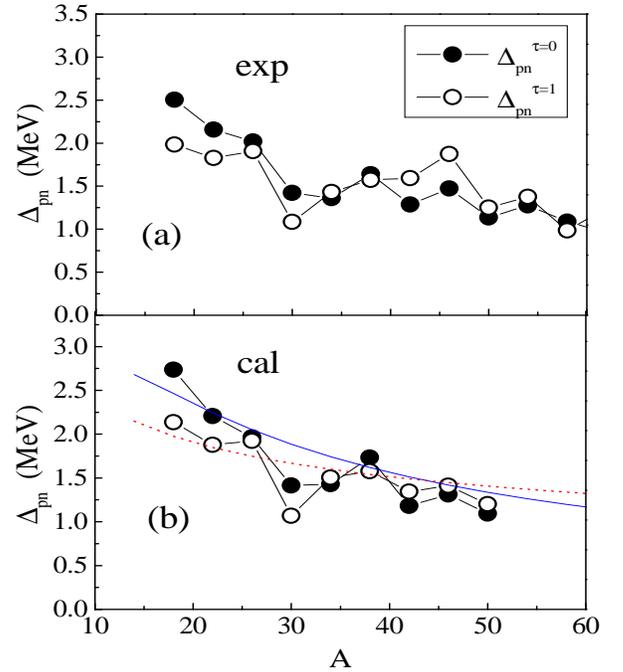}
  \caption{$\tau=0$ and $\tau=1$ {\it pn} pairing gaps estimated from double 
  differences of binding energies for odd-odd $N=Z$ nuclei: (a) experimental ones; 
  (b) those of shell model calculations. The solid and dotted curves show 
  $122.25(1-1.67A^{-1/3})/A$ and $5.18A^{-1/3}$, respectively. 
  }
  \label{fig1}
\end{figure}

It has been recently reported \cite{Schiller,Melby} that the canonical heat capacities 
extracted from observed level densities in $^{162}$Dy, $^{166}$Er and $^{172}$Yb display 
the S shape with a peak around $T\approx 0.5$ MeV, which is interpreted as the breaking of 
like-nucleon $J=0$ pairs because the BCS critical temperature 
corresponds to $T_{c}\approx 0.57\Delta_{n}(T=0)\approx 0.5$ MeV, where the like-nucleon 
pairing gap $\Delta_{n}(T=0)$ is calculated at zero temperature by the BCS theory. 
Thus it seems that the S shape is a signature of pairing transition at the critical temperature. 
For the finite Fermi system like a nucleus, however, since the nuclear radius is much smaller 
than the coherence length of the Cooper pair, statistical fluctuations beyond the mean field 
in the BCS theory become large. The fluctuations smooth out the sharp phase transition, 
and then the like-nucleon pairing gap $\Delta_{n}$ does not quickly become zero 
at the BCS critical temperature but decreases with 
increasing temperature. There are many approaches to treat the fluctuations beyond the mean 
field. The shell model calculation can take into account the large fluctuations beyond the 
mean field. Recently the shell model Monte Carlo (SMMC) calculation \cite{Liu,Alhassid} using 
the $fp + g_{9/2}$ shell has been performed in the even- and odd-mass Fe isotopes.

We recently proposed \cite{kaneko2} ``thermal" odd-even mass difference to estimate the 
like-nucleon pairing energy at finite temperature,
 and showed in the spherical shell model calculations
  that the drastic suppression of like-nucleon pairing energy
  due to finite temperature brings about the S shape in the heat capacity 
arround the temperature $T_{c}\approx 0.57\Delta_{n}(T=0)$ MeV. 
In this rapid communication, we study the {\it pn} pairing energies at finite temperature
in odd-odd $N=Z$ nuclei. 
Does pairing transition due to the breaking of {\it pn} pairs take place
when temperature increasing? 
It is now interesting to investigate thermal behavior of the $pn$ pairing energies 
in $N=Z$ nuclei. 

We start from the double difference of binding energies 
\cite{Janecke3,Jensen,kaneko3} defined as 
\begin{eqnarray}
\Delta_{pn}^{\tau}(Z,N) =  \frac{1}{2}[B(Z,N)^{\tau} - B(Z,N-1) \nonumber \\
       -  B(Z-1,N)+B(Z-1,N-1)], \label{eq:5}
\end{eqnarray}
where $B(Z,N)$ is the binding energy. 
The indicator $\Delta_{pn}^{\tau=1}$ gives the $\tau=1$ $pn$ 
pairing gap in $N=Z$ nuclei. The $\Delta_{pn}^{\tau=0}$ can be regarded as the $\tau=0$ $pn$ 
pairing gap as well. 
Figure 1 (a) shows the $\tau=0$ and $\tau=1$ $pn$ pairing gaps estimated from the double 
differences of experimental binding energies (1) in odd-odd $N=Z$ nuclei with 
$A=18-58$. 
The $\tau=0$ $pn$ energy is somewhat larger than the $\tau=1$ $pn$ energy
in the $sd$ shell nuclei and {\it vice versa} in the $pf$ shell nuclei.
Over a wide range of odd-odd $N=Z$ nuclei, however, basically Fig. 1 shows 
almost the same magnitude of the $\tau=0$ and $\tau=1$ $pn$ pairing gaps. 
  We carried out shell model calculations using isospin-invariant interactions 
  such as the unified $sd$ (USD) interaction \cite{USD} 
  for odd-odd $N=Z$ nuclei in $sd$ shell and the GPFX1 interaction \cite{GPFX1} for $^{42}$Sc, 
  $^{46}$V, and $^{50}$Mn  in $fp$ shell. 
On the mean-field level the ratio between the strengths of {\it pp-}, {\it nn-}, and 
{\it pn-}pair fields is given by the orientation of the pair field. 
The relative strengths of three types of pair fields becomes only definite when isospin 
symmetry is restored. Note that the shell model calculations with isospin invariance show 
$\Delta^{\tau=1}_{pp}=\Delta^{\tau=1}_{nn}=\Delta^{\tau=1}_{pn}$ in odd-odd $N=Z$ nuclei. 

In Fig. 1(b), we can see that the shell model results 
reproduce well the experimental $pn$ pairing gaps, and describe the characteristic behavior 
in Fig. 1(a). 
The $pn$ pairing gaps are closely realted to the energy difference 
$B(Z,N)^{\tau=1}-B(Z,N)^{\tau=0}$ between the lowest 
$\tau=0$ and $\tau=1$ states in odd-odd $N=Z$ nuclei, 
because the energy difference satisfies the following identity \cite{kaneko1}, 
 \begin{eqnarray}
B(Z,N)^{\tau=1}-B(Z,N)^{\tau=0} & = & 2(\Delta_{pn}^{\tau=0}-\Delta_{pn}^{\tau=1}). 
\end{eqnarray}
Odd-odd $N=Z$ nuclei with $A < 40$ have the ground states with $\tau=0, J > 0$ except for 
$^{34}$Cl, while the ground states of odd-odd $N=Z$ nuclei with $40 < A < 74$ are 
$\tau=1$ and $J=0$ except for $^{58}$Cu. 
Several authors \cite{Vogel,Janecke1,Zeldes,Macc,Janecke2,Frauendorf} 
discussed that this degeneracy is attributed to the delicate 
balance between the symmetry energy $a(A)\tau (\tau + 1)/A$ 
and pairing gap $\Delta$ and that the energy difference 
$\delta B=B(Z,N)^{\tau=1}-B(Z,N)^{\tau=0}$ is expressed as $\delta B=2(a(A)/A-\Delta)$. 
However, if we employ the symmetry energy coefficient $a(A)=134.4(1-1.52A^{-1/3})$ 
and pairing gap $\Delta=5.18A^{-1/3}$ of Duflo and Zuker mass formula \cite{Duflo}, 
the energy difference $\delta B$ becomes larger than 
the experimental value. 
As suggested in our previous paper, the isoscalar pairing gap $\Delta^{\tau=0}_{pn}$
 is approximately written as $122.25(1-1.67A^{-1/3})/A$ and the 
isovector one $\Delta^{\tau=1}_{pn}$ is equal to the like-nucleon $nn$ pairing gap
 $\Delta_{n} \approx 5.18A^{-1/3}$. 
These two curves are shown in Fig. 1(b) for comparison. 
Since $\delta B=2(\Delta^{\tau=1}_{pn}-\Delta^{\tau=0}_{pn})$, 
the degeneracy between the lowest 
$\tau=0$ and $\tau=1$ states in odd-odd $N=Z$ nuclei comes from the 
delicate balance between the isoscalar and isovector {\it pn} pairing energies. 

Let us next describe the $pn$ pairing gaps at finite temperature. 
We introduce the canonical partition function defined by 
\begin{eqnarray}
Z(T) = {\rm Tr}({\rm e}^{-H/T}) = \sum_{i=0}^{\infty}{\rm e}^{-E_{i}/T}, 
\label{eq:1}
\end{eqnarray}
where $E_{i}$ is the energy of 
the $i$th eigenstate with degeneracies based on symmetries for the Hamiltonian $H$ of a system. 
All the eigenvalues $E_{i}$ are obtained by solving the eigenvalue equations 
$H\Psi_{i}=E_{i}\Psi_{i}$. 
Then, the partition function in the canonical ensemble is calculated from Eq. (\ref{eq:1}), 
and any thermodynamical quantities $O(T)$ can be evaluated from 
\begin{eqnarray}
O(T) = \langle O \rangle = {\rm Tr}(O{\rm e}^{-H/T})/Z(T), 
\label{eq:2}
\end{eqnarray}
where $\langle O \rangle$ stands for the average value of operator $O$ over the range of 
eigenstates. For instance, the thermal energy is expressed as
\begin{eqnarray}
E(Z,N,T) = \langle H \rangle = \sum_{i=0}^{\infty}E_{i}{\rm e}^{-E_{i}/T}/Z(T). 
\label{eq:3}
\end{eqnarray}
The heat capacity is then given by 
\begin{eqnarray}
C(Z,N,T) = \frac{\partial E(Z,N,T)}{\partial T}.
\label{eq:4}
\end{eqnarray}

We now introduce the following double difference of ``thermal" enegies
$E(Z,N,T)$ analogous to Eq. (1) as an indicator of $pn$ pairing energies, 
\begin{eqnarray}
\Delta_{pn}^{\tau}(Z,N,T) =  \frac{1}{2}[E(Z,N,T)^{\tau} - E(Z,N-1,T) \nonumber \\
       -  E(Z-1,N,T)+E(Z-1,N-1,T)]. \label{eq:5}
\end{eqnarray}
The double differences of binding energies at zero temperature in Eq. (1) 
are known theoretically and experimentally as important quantities in evaluation of the {\it pn} 
pairing energies in a nucleus \cite{Janecke3,Jensen,kaneko3}. 
 The double differences of thermal energies in Eq. (7) are also 
indicators of the $pn$ pairing energies and can be regarded as the $pn$ pairing gaps 
at finite temperature. 

Let us evaluate the double difference of thermal energies 
(7) for $N=Z$ $sd$ shell nuclei. We make numerical calculations 
by way of two steps. First, we carry out the exact shell model calculations
in the $sd$ shell using the USD interaction \cite{USD},
 and calculate the correlated thermal energy $E_{v,tr}$
 from Eq. (\ref{eq:3}). 
Secondly, we extend the model space to a larger one
($sd + fp + s_{1/2}d_{5/2}$) in order to display the double difference
of thermal energies in a broader range of temperature 
using an independent-particle approximation
\cite{Alhassid}. The single-particle energies of the extended space
are obtained by diagonalizing the Woods-Saxon potential with the spin-orbit
interaction, where the harmonic-oscillator (H.O.) eigenfunctions
are used. The Woods-Saxon parameters are chosen so as to reproduce
the single-particle energies estimated from $^{17}$O, because it is
necessary to reasonably extrapolate the single-particle energies of
the $sd$ shell to those of the larger space.
In this way, we combine the correlated thermal energy $E_{v,tr}$ 
in the truncated space with the thermal energy $E_{sp}$ calculated
using the independent-particle approximation in the larger space.
The thermal energy which takes account of the interaction effects
in the sd shell is estimated as follows \cite{Alhassid}:
\begin{eqnarray}
E & = & E_{v,tr} + E_{sp} - E_{sp,tr},
\label{eq:6}
\end{eqnarray}
where $E_{sp,tr}$ is the thermal energy of the $sd$ shell within the
independent-particle approximation.
We now obtain the double difference of thermal energies $\Delta_{pn}^{\tau}$
by substituting $E$ of Eq. (\ref{eq:6}) for $E(Z,N,T)$ in Eq. (7). 

\begin{figure}[t]
\includegraphics[width=8cm,height=10cm]{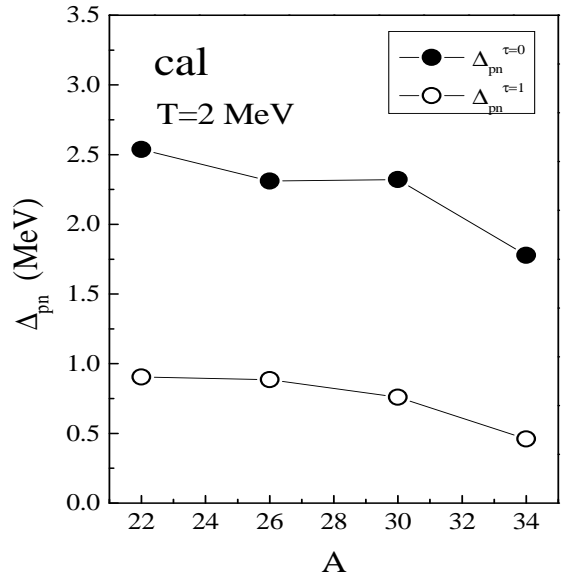}
  \caption{Calculated thermal $pn$ pairing gaps for odd-odd $N=Z$ nuclei at temperature 
  $T=$ 2.0 MeV. The solid circles denote the $\tau$=0 {\it pn} pairing gap, and the 
  open circles the $\tau$=1 {\it pn} pairing gap.}
  \label{fig2}
\end{figure}

\begin{figure}[t]
\includegraphics[width=8cm,height=10cm]{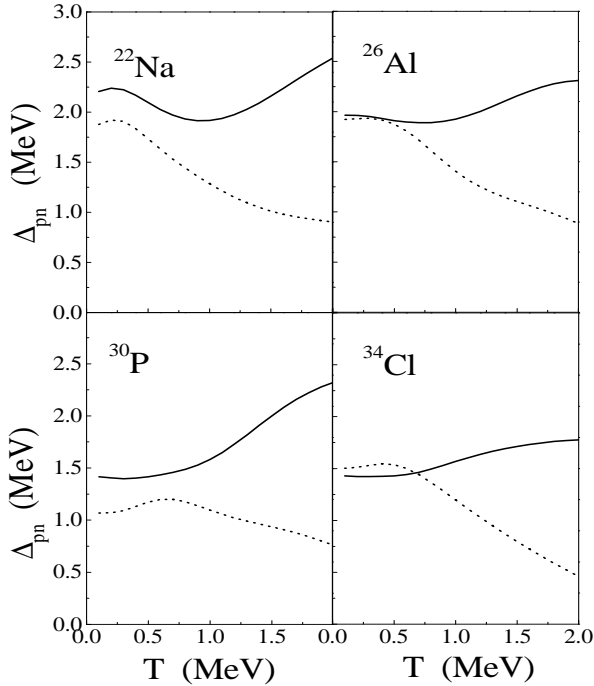}
  \caption{Calculated thermal $pn$ pairing gaps for odd-odd $N=Z$ nuclei as a function of 
  temperature. The solid line denotes the $\tau$=0 {\it pn} pairing gap, and the 
  dotted line the $\tau$=1 {\it pn} pairing gap.}
  \label{fig3}
\end{figure}
\begin{figure}[t]
\includegraphics[width=8cm,height=10cm]{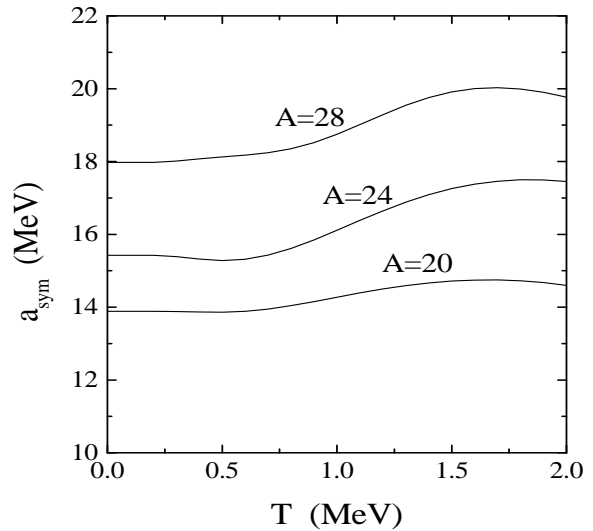}
  \caption{Symmetry energy coefficient $a_{sym}$ as a function of temperature for 
           $N\approx Z$ nuclei with mass number $A=$20, 24, and 28.}
  \label{fig4}
\end{figure}

Figure 2 shows the calculated thermal $pn$ pairing gaps for odd-odd $N=Z$ nuclei, 
$^{22}$Na, $^{26}$Al, $^{30}$P, and $^{34}$Cl at temperature $T=$ 2.0 MeV. 
 The $\tau=$1 and $\tau=$0 $pn$ pairing gaps are largely 
separated at $T=$ 2.0 MeV.
Comparing Fig. 2 with Fig. 1(b), we notice that the $\tau=$1 $pn$ pairing gap
decreases but the $\tau=$0 $pn$ pairing gap keeps the magnitude
from zero temperature to high temperature. 

Figure 3 shows the variation of the thermal $pn$ pairing gaps 
depending on temperature $T$
for $^{22}$Na, $^{26}$Al, $^{30}$P, and $^{34}$Cl. 
In all graphs, we can see increase of the $\tau=$0 $pn$ pairing gap and decrease of the 
$\tau=$1 $pn$ pairing gap. 
As mentioned above, at zero temperature the $\tau=$0 and $\tau=$1 $pn$ pairing gaps 
are almost the same, and the lowest $\tau=$0 and $\tau=$1 states are degenerate. 
As increasing temperature, the $\tau=$1 $pn$ pairing gap decreases and $\tau=$0 
one rather increases. 
Thus, we know that the $\tau=0$ pairing energy becomes dominant at high temperature. 

It would be valuable to discuss the symmetry energy $\sim 4a_{\rm sym}(T)\tau (\tau + 1)/A$ 
at finite temperature because it is closely related to the $\tau=0$ pairing 
energy. In our previous paper \cite{kaneko3}, we suggested that the dominant part of the 
symmetry energy comes from the $\tau=0$ pairing energy part in the shell-model interaction energy. 
For the application of the symmetry energy in core-collapse supernova simulations, Donati 
{\it et al.} \cite{Donati} pointed out a possibility that the symmetry energy coefficient 
$a_{\rm sym}$ at the finite temperature has been estimated to be somewhat larger than that of 
stable nuclei at zero temperature. 
The increase $(\sim 3\%)$ of the symmetry energy between 
$T=0.0$ and $T=1.0$ MeV after implementing the correction in the SMMC 
calculations is smaller than that $(\sim 8\%)$ of the quasiparticle random phase 
approximation (QRPA) with temperature \cite{Dean}. 
We now calculate the temperature dependence of the symmetry energy using the shell model 
calculations. 
We estimate the symmetry energy coefficient from the thermal energy $E(Z,N,T)^{\tau}$ 
with isospin $\tau$ and temperature $T$ obtained in the shell model calculations as follows: 
\begin{eqnarray}
a_{\rm sym}(T) & = & \frac{E(Z,N,T)^{\tau}-E(Z,N,T)^{\tau'}}{\tau (\tau + 1) - \tau' (\tau' + 1) }A, 
\label{eq:7}
\end{eqnarray}
where $\tau$ and $\tau'$ are different isospins for isobaric nuclei with same mass 
number $A$. 
At zero temperature, the calculated symmetry energy coefficient $a_{\rm sym}(T=0)\sim 16$ MeV 
for $A$=24 is in good agreement with the value determined from experimental masses and 
with the empirical value of Duflo and Zuker mass formula. 

Figure 4 shows the symmetry energy coefficient $a_{\rm sym}$ as a function of the temperature 
for even-even $N\approx Z$ nuclei with mass number $A=$20, 24, and 28, 
where several isobaric pairs of $N\approx Z$ nuclei such as $(^{20}$Ne,$^{20}$O), 
$(^{24}$Mg,$^{24}$Ne), and $(^{28}$Si,$^{28}$Mg) are chosen. 
This figure shows that the symmetry energy coefficients increase 
with increasing temperature in these three cases. 
Moreover, we can see that the symmetry energy coefficient depends on the mass $A$ 
which is empirically fitted by adding the surface contribution 
with the $A^{-1/3}$ dependence at zero temperature. 
This mass dependence appears in the $\tau=0$ {\it pn} pairing gap estimated from 
the double difference of binding energies, in Fig. 1(b). 
Figure \ref{fig4} also suggests that the mass dependence changes 
as temperature increases. 
To see the temperature dependence of the symmetry energy coefficient, we define the 
relative change of the symmetry energy coefficient with respect to temperature as 
\begin{eqnarray}
\delta a_{\rm sym}(T) & = & \frac{a_{\rm sym}(T) - a_{\rm sym}(T=0)}{a_{\rm sym}(T=0)}. 
\label{eq:8}
\end{eqnarray}
Averaging the $\delta a_{\rm sym}(T)$ at $T=1.0$ MeV over various pairs of nuclei, 
we obtain an increase $\sim 4\%$. This is in agreement with the SMMC result $\sim 3\%$ 
obtained after implementing the correction. 
We used here the form of symmetry energy $\tau(\tau+1)$,
which is motivated by the charge independence of the nuclear force.
But as a phenomenological parametrization the isospin dependence $\tau(\tau+\alpha)$ 
with $\alpha\neq 1$ is also possible, where the linear term in $\tau$
is so-called Wigner term. 
Recently, empirical fitting to the Wigner term gave $\alpha=1.25$ in 
the vicinity of the $N=Z$ line \cite{Janecke1, Satula}. 
However, the symmetry energy coefficient is affected little by replacing $\tau (\tau + 1)$ 
with $\tau ( \tau + 1.25 )$. Moreover, by definition the relative change of the symmetry 
energy coefficient $\delta a_{\rm sym}(T)$ does not change by this replacement. 

In conclusion, we investigated the $\tau$=0 and $\tau$=1 {\it pn} pairing energies 
at finite temperature using the shell model calculations. 
The {\it pn} pairing gaps at finite temperature were estimated from the double differences 
of thermal energies defined by Eq. (7),
 which is analogous to the double differences of binding 
energies as indicators of the {\it pn} pairing energies at zero temperature. 
It was shown that as temperature increases the isoscalar {\it pn} pairing 
energy rather increases, while the isovector {\it pn} pairing energy decreases. 
Almost the same {\it pn} pairing gaps of stable $N=Z$ nuclei at zero temperature are 
separated with increasing temperature. 
We also studied the temperature dependence of the symmetry energy in $N\approx Z$ nuclei. 
The symmetry energy coefficients increase with increasing temperature.
 The increase of the calculated 
symmtery energy coefficient between $T=0.0$ and $T=1.0$ MeV is in good agreement with that 
of the SMMC calculations. We suggest that the {\it pn} pairing energies can be estimated 
using Eqs. (5) and (7) from the measured level densities of nuclei. 
We expect that the {\it pn} pairing energies play an important role in the 
astrophysics.



\end{document}